\documentclass{appolb}
\usepackage{amsmath}
\usepackage{cite}
\usepackage{graphicx}

\newcommand{\colorfulmethod}	{CoLorFulNNLO}
\newcommand{\mccsm}	{{\texttt{MCCSM}}}

\newcommand{\amcatnlo}	{\texttt{MadGraph5\_aMC@NLO}}
\newcommand{\eerad}	{\texttt{EERAD3}}

\newcommand{\gev}	{\ensuremath{\,\mathrm{GeV}}}

\newcommand{\as}	{\ensuremath{\alpha_{\rm s}}}

\newcommand\eqn[1]     {Eq.\,(\ref{#1})}
\newcommand\eqns[2]    {Eqs.\,(\ref{#1}) and~(\ref{#2})}

\newcommand\fig[1]     {Fig.\,\ref{#1}}

\begin{document}

%%%%%%%%%%%%%%%%%%%%%%%%%%%%%%%%%%%%%%%%%%%%%%%%%%%%%%%%%%%%
% Front matter                                             %
%%%%%%%%%%%%%%%%%%%%%%%%%%%%%%%%%%%%%%%%%%%%%%%%%%%%%%%%%%%%

\title{Higher order corrections in the CoLoRFulNNLO framework%
\thanks{Presented at the XXIII Cracow Epiphany Conference}%
}
\author{G\'abor Somogyi\footnote{Speaker}, Adam Kardos, Zolt\'an Sz\H or and 
\\
Zolt\'an Tr\'ocs\'anyi
\address{University of Debrecen and MTA-DE Particle Physics Research Group \\
H-4010 Debrecen, PO Box 105, Hungary}
}

\maketitle

%%%%%%%%%%%%%%%%%%%%%%%%%%%%%%%%%%%%%%%%%%%%%%%%%%%%%%%%%%%%
% Abstract                                                 %
%%%%%%%%%%%%%%%%%%%%%%%%%%%%%%%%%%%%%%%%%%%%%%%%%%%%%%%%%%%%

\begin{abstract}
We discuss the \colorfulmethod\ method for computing higher order radiative corrections 
to jet cross sections in perturbative QCD. We apply our method to the calculation of 
event shapes and jet rates in three-jet production in electron-positron annihilation. 
We validate our code by comparing our predictions to previous results in the literature
and present the jet cone energy fraction distribution at NNLO accuracy. 
We also present preliminary NNLO results for the three-jet rate using the Durham jet 
clustering algorithm matched to resummed predictions at NLL accuracy, and a comparison to 
LEP data. 
\end{abstract}

% 13.66.Bc Hadron production in e?e+ interactions
% 13.87.-a Jets in large-Q2 scattering 

%%%%%%%%%%%%%%%%%%%%%%%%%%%%%%%%%%%%%%%%%%%%%%%%%%%%%%%%%%%%
% Introduction                                             %
%%%%%%%%%%%%%%%%%%%%%%%%%%%%%%%%%%%%%%%%%%%%%%%%%%%%%%%%%%%%

\section{Introduction}

The strong coupling \as\ is one of the fundamental parameters of the 
standard model of particle physics thus its precise determination is mandatory. 
Nowadays, event shapes and jet rates measured in three-jet formation in 
electron-positron annihilation are still among the most precise tools used 
for accurate extractions of \as\ from data. In these analyses, the measurement 
of \as\ involves fitting theoretical predictions for a given observable and 
collider energy to observations. Hence theoretical input is essential and the 
goodness of the fitting procedure relies heavily on the quality of the theoretical 
predictions used.

In high-energy particle physics we may approach the quality of theoretical 
predictions from two perspectives. First, the general framework for performing 
calculations is perturbation theory, where in case of QCD the perturbative 
parameter is the strong coupling, \as. Due to the complexity of calculations 
only the first few terms of the perturbative series can be evaluated and this 
truncation introduces a theoretical uncertainty manifested by the dependence of 
the predictions on non-physical parameters such as the renormalization and factorization 
scales. Thus one way of increasing the theoretical precision of the calculations is by 
including exact higher order corrections in perturbation theory. Second, the actual 
calculations of physical observables involve numerical integrations over the physical 
phase space and this introduces a statistical uncertainty in the predictions. Hence 
formal higher order precision must be supplemented by good numerical accuracy to obtain 
results that are useful for experimental needs.

For three-jet production in electron-positron annihilation, all matrix elements 
necessary for the computation of next-to-next-to-leading order (NNLO) corrections 
have been known in the literature for some time \cite{Bern:1996ka,Bern:1997sc,Garland:2001tf,Garland:2002ak} and indeed NNLO corrections to several event shapes 
\cite{GehrmannDeRidder:2007hr,Weinzierl:2009ms,DelDuca:2016csb,DelDuca:2016ily}
and jet rates have been evaluated  \cite{GehrmannDeRidder:2008ug,Weinzierl:2010cw}. 
Hence this process is not only interesting from a 
phenomenological point of view, but it also provides an ideal testing ground for 
new computational methods at this order in perturbation theory. In this contribution we 
summarize a completely local subtraction method, called \colorfulmethod, for computing 
QCD corrections to jet cross sections at NNLO accuracy and present the application of our 
framework to three-jet production in electron-positron annihilation. Our method 
demonstrates excellent numerical stability and accuracy for the considered observables.

%%%%%%%%%%%%%%%%%%%%%%%%%%%%%%%%%%%%%%%%%%%%%%%%%%%%%%%%%%%%
% The CoLoRFulNNLO method                                  %
%%%%%%%%%%%%%%%%%%%%%%%%%%%%%%%%%%%%%%%%%%%%%%%%%%%%%%%%%%%%

\section{The \colorfulmethod\ method}

In perturbative QCD the expansion of a jet cross section defined by some physical
quantity $J$ can be formally written up to NNLO accuracy as
\begin{align}
\sigma[J] &= \sigma^{\rm LO}[J] + \sigma^{\rm NLO}[J] + \sigma^{\rm NNLO}[J] + \dots
\,.
\end{align}
Focusing on the production of $m$ jets from a colorless initial state, the leading order 
(LO) cross section is simply given by integrating the fully differential Born cross section 
for the production of $m$ partons over the $m$-parton phase space defined by the observable $J$,
\begin{align}
\sigma^{\rm LO}[J] &= \int_m{\rm d}\sigma^{\rm B}_{m}J_{m}
\,.
\end{align}
The next-to-leading order (NLO) correction can be written as the sum of two terms,
\begin{align}
\sigma^{\rm NLO}[J] &= \int_{m+1}{\rm d}\sigma^{\rm R}_{m+1}J_{m+1} 
+ \int_{m}{\rm d}\sigma^{\rm V}_{m}J_{m}
\,,
\end{align}
and is finite for any infrared-safe observable by the general theorem of Kinoshita, Lee 
and Nauenberg. Although the KLN theorem guarantees the finiteness for the sum of the real 
emission ($\sigma^{\rm R}_{m+1}$) and virtual ($\sigma^{\rm V}_{m}$) corrections, it does 
not say anything about the contributions separately which are indeed infinite in four 
spacetime dimensions. Using conventional dimensional regularization in $d = 4 - 2\epsilon$ 
dimensions to regularize the two pieces, the singularities become poles in $\epsilon$, 
which nevertheless cancel between the two contributions in the final result. 
However, this cancellation is not manifest. On one hand the singularities in the real 
emission part have a kinematic origin: they are due to divergent phase space integrals when 
one final state parton becomes unresolved. On the other hand, the $\epsilon$-poles in the 
virtual correction arise from the integration over the loop momentum. In general the squared 
matrix elements and observables in QCD are much too complicated to perform an analytic 
calculation in $d$ dimensions, and our aim is to carry out the computations in four dimensions 
using Monte Carlo integration techniques. To do so, the real and virtual emission contributions 
have to be made finite separately which we achieve by local subtractions. In this method, 
an approximate differential cross section, ${\rm d}\sigma^{\rm R,A_1}_{m+1}$, is subtracted 
from the real emission contribution. This approximate cross section is constructed carefully 
to have the same kinematic singularity structure (in $d$ dimensions) as the real emission 
cross section. Thus, the difference is free of non-integrable kinematic singularities and 
the phase space integral can be evaluated in four dimensions using standard Monte Carlo 
techniques. The poles appearing in the virtual contribution are then removed by adding back 
the approximate cross section after integrating over the momentum and summing over the 
quantum numbers (color, flavor) of the unresolved particle (these operations are collectively 
denoted by $\int_1$). Then the NLO correction takes the form
\begin{align}
\sigma^{\rm NLO}[J] =&
\int_{m+1}\left[{\rm d}\sigma^{\rm R}_{m+1}J_{m+1} - {\rm d}\sigma^{\rm R,A_1}_{m+1}J_{m}\right]_{d=4}
\notag\\ 
&+ \int_{m}\left[{\rm d}\sigma^{\rm V}_{m}J_{m}
+ \int_1{\rm d}\sigma^{\rm R,A_1}_{m+1}J_{m}\right]_{d=4}\,,
\end{align} 
where now both contributions are finite as discussed. Several explicit constructions 
are available for the approximate cross section ${\rm d}\sigma^{\rm R,A_1}_{m+1}$ in 
the literature \cite{Frixione:1995ms,Catani:1996vz,Nagy:1996bz,Somogyi:2006cz,Somogyi:2009ri}.

The NNLO correction is composed of three different contributions,
\begin{align}
\sigma^{\rm NNLO}[J] = 
\int_{m+2}{\rm d}\sigma_{m+2}^{\rm RR}J_{m+2}
+ \int_{m+1}{\rm d}\sigma_{m+1}^{\rm RV}J_{m+1}
+ \int_{m}{\rm d}\sigma_{m}^{\rm VV}J_{m}
\,.
\end{align}
The first term is the double real (RR) piece which involves tree level squared 
matrix elements with $m+2$-parton kinematics and develops singularities when one or 
two partons become unresolved. The second term is the real-virtual one (RV) and contains 
the interference of one-loop and tree level matrix elements with $m+1$-parton kinematics. 
This contribution develops both kinematic singularities when a parton becomes unresolved 
and also contains explicit $\epsilon$-poles coming from one-loop amplitudes. Finally, the 
third term is the double virtual (VV) contribution, which includes the interference of 
the $m$-parton two-loop and tree level matrix elements as well as the square of the 
$m$-parton one-loop matrix element. This contribution is free from kinematic singularities 
(the infrared-safe jet function screens any remaining divergences of the squared matrix 
elements), but it contains explicit 
$\epsilon$-poles which come from integrations over loop momenta.

The idea behind the \colorfulmethod\ method is to define completely local subtraction 
terms for the NNLO correction in the same spirit as was done at NLO accuracy. Thus the 
$m+2$-parton contribution is made finite by introducing local subtraction terms whose 
kinematic singularities exactly reproduce those of the double real emission matrix 
elements (in $d$ dimensions) in each single and double unresolved limit:
\begin{align}
\sigma^{\rm NNLO}_{m+2}[J] =
\int_{m+2}\bigg\{&
{\rm d}\sigma_{m+2}^{\rm RR}J_{m+2}
- {\rm d}\sigma_{m+2}^{\rm RR,A_2}J_{m}
\notag\\&
- \bigg[{\rm d}\sigma_{m+2}^{\rm RR,A_1}J_{m+1}
- {\rm d}\sigma_{m+2}^{\rm RR,A_{12}}J_{m}
\bigg]
\bigg\}_{d=4}
\,.
\label{eq:sigNNLOm+2}
\end{align}
In \eqn{eq:sigNNLOm+2}, ${\rm d}\sigma^{\rm RR,A_2}_{m+2}$ regularizes those 
singularities of the RR contribution which emerge in double unresolved limits, 
while ${\rm d}\sigma^{\rm RR,A_1}_{m+2}$ serves as a local counterterm for single 
unresolved singularities. The last term, ${\rm d}\sigma^{\rm RR,A_{12}}_{m+2}$, 
is introduced to remove both the kinematic singularities that develop in 
${\rm d}\sigma^{\rm RR,A_2}_{m+2}$ in single unresolved regions and also the 
singularities of ${\rm d}\sigma^{\rm RR,A_1}_{m+2}$ in double unresolved ones. 
The precise definitions of all subtraction terms that appear in \eqn{eq:sigNNLOm+2} 
were given in \cite{Somogyi:2006da}.

The $m+1$-parton contribution takes the form
\begin{align}
\sigma^{\rm NNLO}_{m+1} =
\int_{m+1}\bigg\{&
\bigg(
{\rm d}\sigma_{m+1}^{\rm RV}
+ \int_1{\rm d}\sigma^{\rm RR,A_1}_{m+2}
\bigg)J_{m+1}
\notag \\&
- \bigg[
{\rm d}\sigma_{m+1}^{\rm RV,A_1}
+ \left(\int_1{\rm d}\sigma_{m+2}^{\rm RR,A_1}\right)\!\strut^{\rm A_1}
\bigg]J_{m}
\bigg\}_{d=4}
\,,
\label{eq:sigNNLOm+1}
\end{align}
where the first line of \eqn{eq:sigNNLOm+1} contains the RV contribution as 
well as the integrated form of the single unresolved subtraction term in \eqn{eq:sigNNLOm+2}, 
$\int_1{\rm d}\sigma^{\rm RR,A_1}_{m+2}$. The sum of these two terms is free of 
$\epsilon$-poles \cite{Somogyi:2006cz}, however both terms still contain kinematical 
singularities when a parton becomes unresolved. These singularities are regularized 
by the local subtraction terms on the second line of \eqn{eq:sigNNLOm+1}. The exact 
definitions of these subtraction terms were presented in \cite{Somogyi:2006db}.

The last contribution to the NNLO correction is the $m$-parton one which contains 
the VV contribution along with the integrated forms of all remaining subtraction 
terms which we have not yet added back. Schematically this can be written as
\begin{align}
\sigma^{\rm NNLO}_{m} =
\int_{m}\bigg\{&
{\rm d}\sigma_{m}^{\rm VV}
+ \int_{2}\bigg[{\rm d}\sigma^{\rm RR,A_2}_{m+2} 
- \sigma^{\rm RR,A_{12}}_{m+2}\bigg]
\notag\\&
+ \int_1\bigg[
{\rm d}\sigma^{\rm RV,A_1}_{m+1}
+ \bigg(\int_1{\rm d}\sigma^{\rm RR,A_1}_{m+2}\bigg)\!\strut^{\rm A_1}
\bigg]
\bigg\}_{d=4}J_{m}
\,.
\label{eq:sigNNLOm}
\end{align} 
Since the $m+2$- and $m+1$-parton contributions in \eqns{eq:sigNNLOm+2}{eq:sigNNLOm+1} 
are both finite by construction, the finiteness of the $m$-parton piece in 
\eqn{eq:sigNNLOm} is automatic and guaranteed by the KLN theorem. The various integrated 
approximate cross sections that appear in \eqn{eq:sigNNLOm} above were computed in a series 
of papers \cite{Somogyi:2008fc}, culminating in the explicit demonstration of the finiteness 
of this contribution for electron-positron annihilation into three jets in 
\cite{DelDuca:2016ily}.
Since the subtractions render all three contributions finite, they can be separately 
integrated numerically using standard Monte Carlo techniques. We stress that since all 
subtractions are completely local, the integrations may be performed with any convenient 
numerical procedure.

%%%%%%%%%%%%%%%%%%%%%%%%%%%%%%%%%%%%%%%%%%%%%%%%%%%%%%%%%%%%
% Electron-positron annihilation into three jets           %
%%%%%%%%%%%%%%%%%%%%%%%%%%%%%%%%%%%%%%%%%%%%%%%%%%%%%%%%%%%%

\section{Electron-positron annihilation into three jets}

We have implemented the \colorfulmethod\ scheme as outlined above into the 
\texttt{fortran90} program library \mccsm\ (Monte Carlo for the \colorfulmethod\ 
Subtraction Method). The implementation is completely general for processes with 
colorless initial states, with only the squared matrix elements for a given process 
(including the color- and spin-correlated ones) as necessary inputs. 

As a first application, we used our method and code to compute NNLO QCD corrections to 
physical observables in three-jet production in electron-positron annihilation 
\cite{DelDuca:2016csb,DelDuca:2016ily}. 
Since these corrections are known for several quantities in the literature 
\cite{GehrmannDeRidder:2007hr,GehrmannDeRidder:2008ug,Weinzierl:2009ms,Weinzierl:2010cw}, 
(see also \cite{Ridder:2014wza} which describes the \eerad\ program implementing the computations of 
\cite{GehrmannDeRidder:2007hr,GehrmannDeRidder:2008ug}) this provides an excellent opportunity 
to validate our method and the framework implementing it. Hence, we compared our predictions 
for the six standard event shape variables of thrust ($T$), C-parameter, total- and wide-jet 
broadening, heavy-jet mass and the two-to-three jet transition variable $y_{23}$ in the 
Durham jet clustering algorithm to the predictions of \cite{GehrmannDeRidder:2007hr,
Weinzierl:2009ms}. We performed the comparisons 
at the LEP2 energy of $\sqrt{s} = m_Z = 91.2\gev$. The perturbative coefficients were 
defined using the normalization common at lepton-lepton colliders:
\begin{align}
\frac{O}{\sigma_0}\frac{{\rm d}\sigma}{{\rm d}O} =
\frac{\alpha_{\rm s}}{2\pi}O\,A(O)
+ \left(\frac{\alpha_{\rm s}}{2\pi}\right)^2 O\,B(O)
+ \left(\frac{\alpha_{\rm s}}{2\pi}\right)^3 O\,C(O)
+ \mathcal{O}(\alpha_{\rm s}^4)
\,,
\end{align}
where $\sigma_0$ is the leading order cross section for $e^+\,e^-\to\,{\rm hadrons}$ and $O$ 
is the event shape variable for which we obtain the NNLO accurate prediction.

\begin{figure}
\centering
\begin{tabular}{cc}
\includegraphics[width=0.47\textwidth]{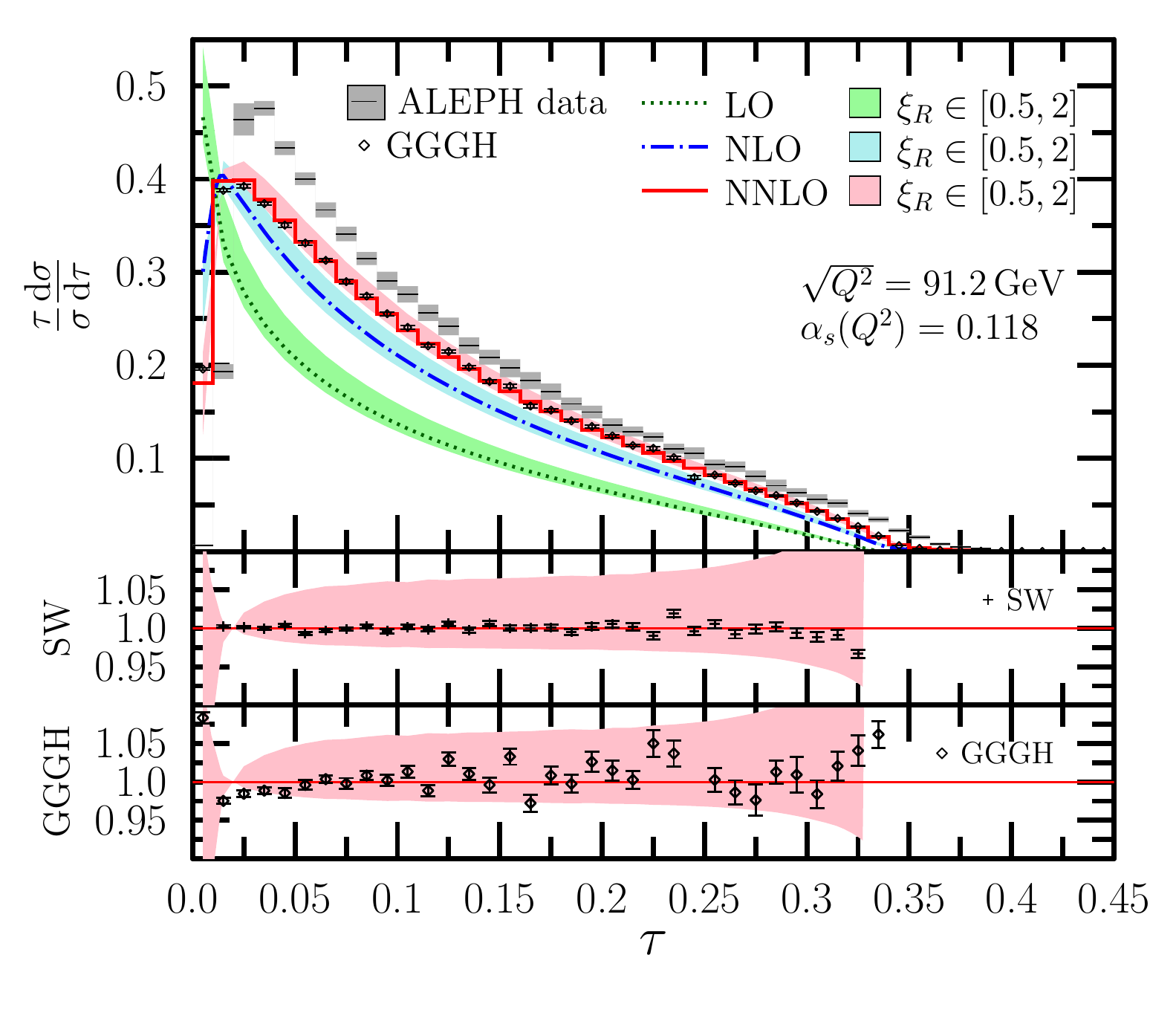} &
\includegraphics[width=0.47\textwidth]{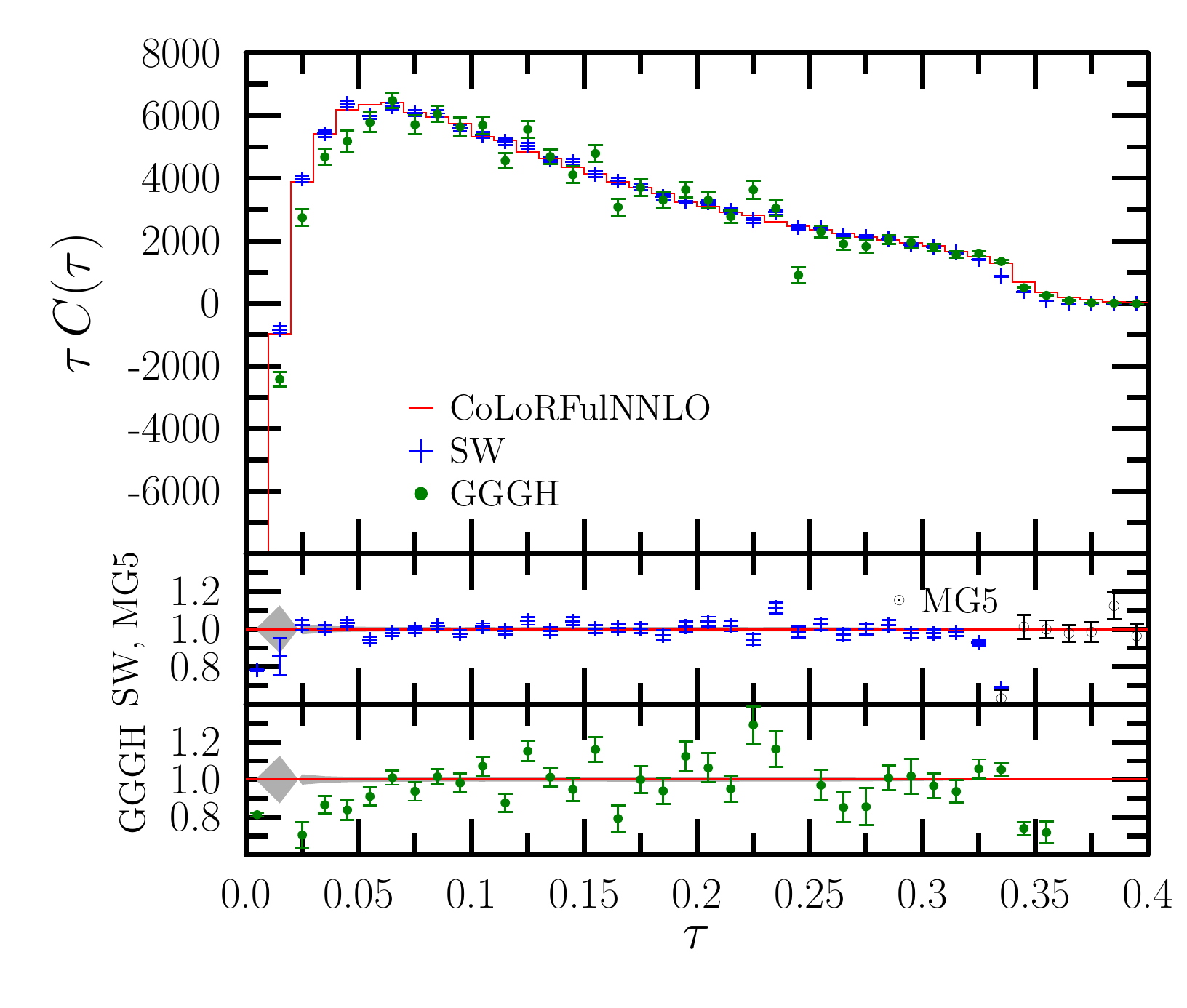} \\
\end{tabular}
\caption{{\label{fig:omTphysCcoeff}} 
Left: physical predictions for thrust ($\tau=1-T$) at 
LO, NLO and NNLO accuracy in QCD with bands representing scale uncertainty. 
Data measured by the ALEPH collaboration \cite{Heister:2003aj} is also shown.
Right: the $\tau\, C(\tau)$ NNLO coefficient of the thrust distribution. On both 
figures the lower panels show the ratio of the predictions of Ref.~\cite{Weinzierl:2009ms} 
(SW) and \eerad\ (GGGH) to \colorfulmethod. In the middle panel of the right figure, results 
from \amcatnlo\ \cite{Alwall:2014hca} (MG5) are also shown above the Born kinematic limit 
of $\tau > 1/3$.
}
\end{figure}

\begin{figure}
\centering
\begin{tabular}{cc}
\includegraphics[width=0.47\textwidth]{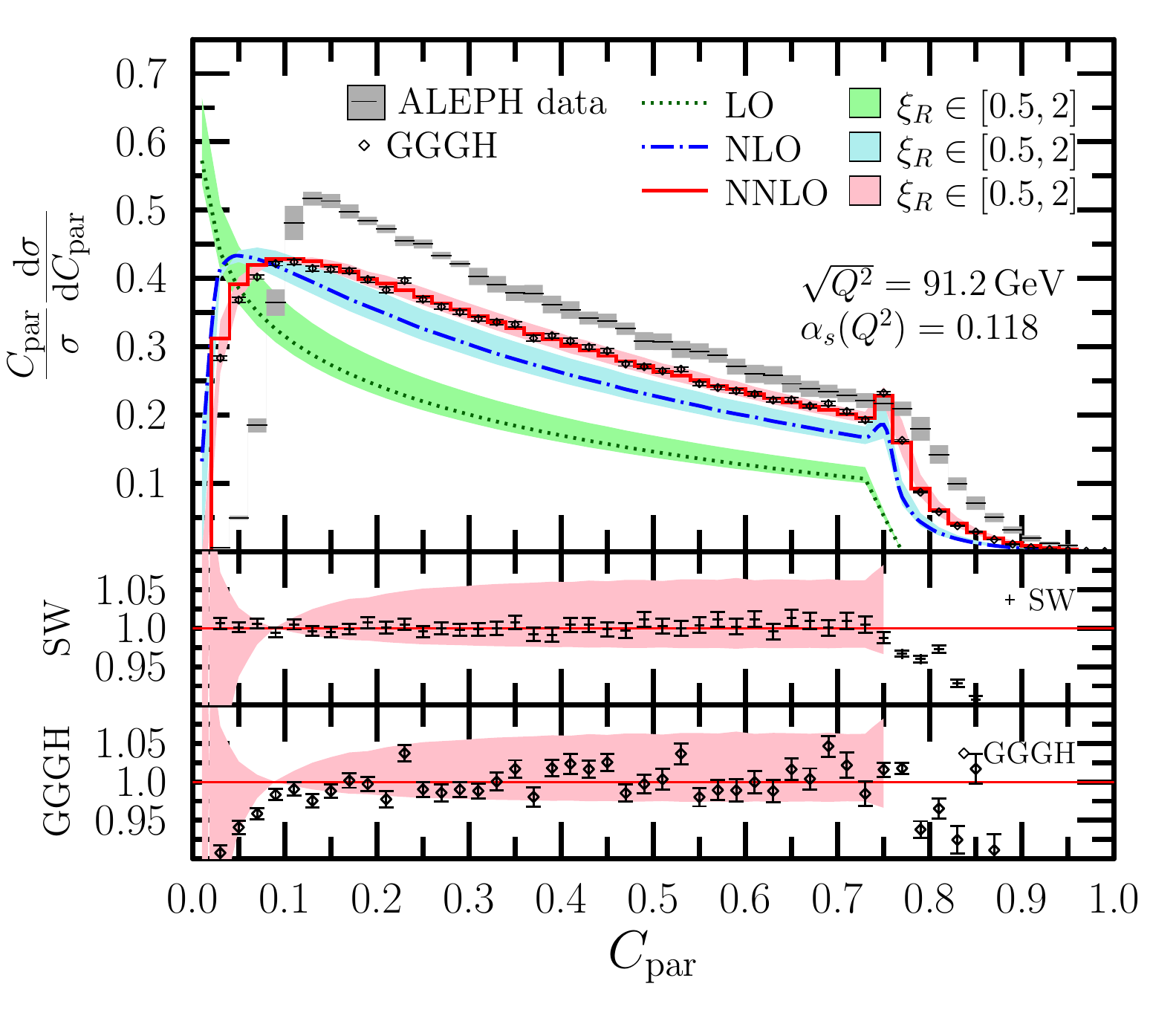} &
\includegraphics[width=0.47\textwidth]{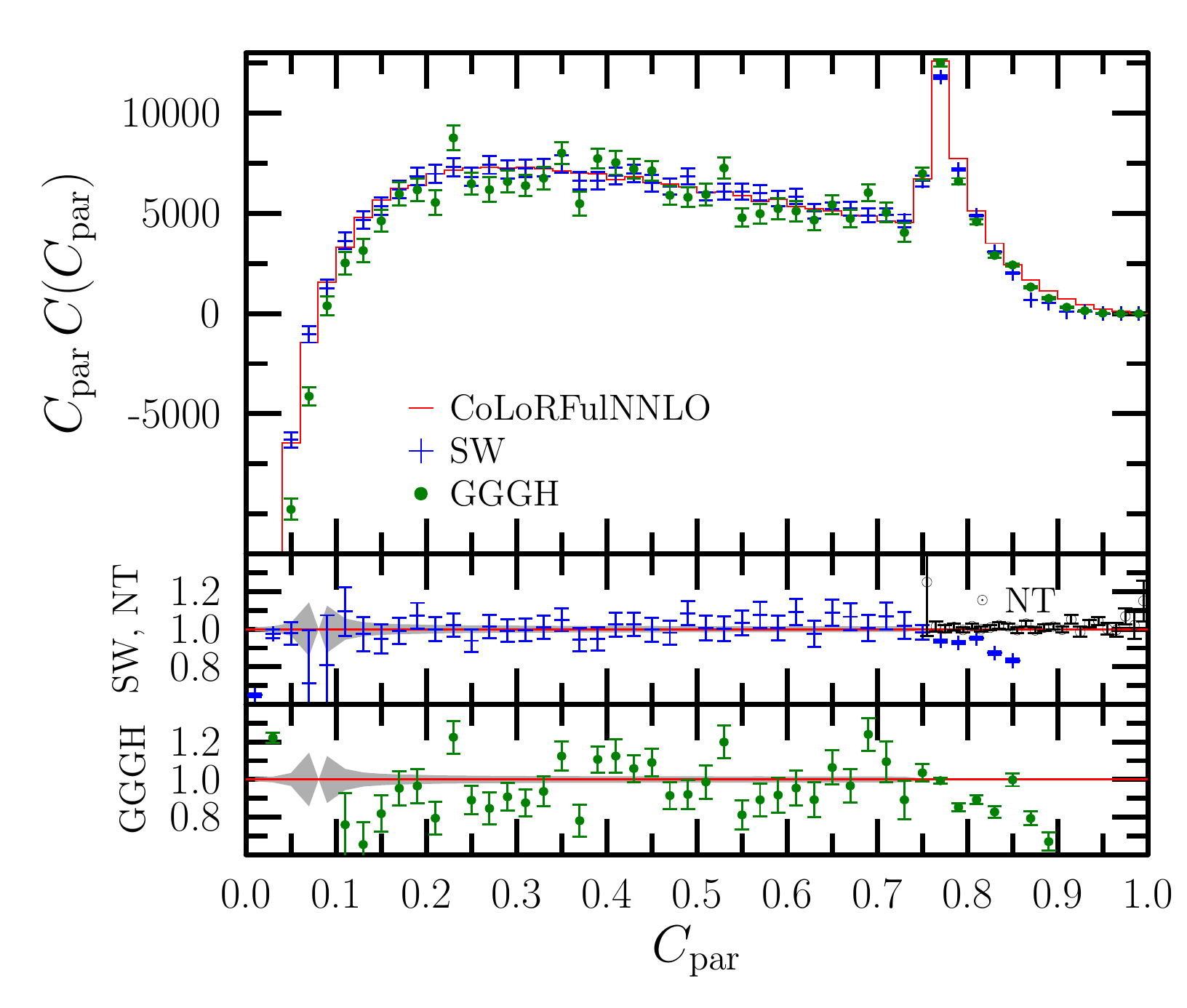} \\
\end{tabular}
\caption{{\label{fig:CparphysCcoeff}} The same as \fig{fig:omTphysCcoeff} for the 
C-parameter distribution. In the middle panel of the right figure, results from 
Ref.~\cite{Nagy:1997yn} (NT) are also shown above the Born kinematic limit of $C_{\mathrm{par}} > 3/4$.}
\end{figure}

We present the comparison of our predictions to those of \eerad\ (denoted by GGGH\footnote{We 
are grateful to G.~Heinrich for providing the predictions of \eerad\ for us.}) and 
Ref.~\cite{Weinzierl:2009ms} (denoted SW\footnote{In these comparisons we use updated 
(with respect to those published in Ref.~\cite{Weinzierl:2009ms}) but unpublished 
predictions provided to us by S.~Weinzierl. We are grateful to S.~Weinzierl for providing 
these updated results for us.}) for two representative cases, thrust ($\tau = 1-T$) 
and the C-parameter in \fig{fig:omTphysCcoeff} and \fig{fig:CparphysCcoeff}. In these plots, 
the left hand panels show the physical predictions for the observables at LO, NLO and NNLO 
accuracy, together with the data measured by the ALEPH collaboration. The bands correspond 
to scale variations in the range of $[m_Z/2,2m_Z]$ with $\mu_0 = m_Z$ chosen as the central 
scale. While these plots clearly show the convergence of the perturbative series for both 
the $\tau$ and C-parameter distributions as we go from LO to NLO and NNLO accuracy, the 
comparison with data also makes it evident that parton shower and non-perturbative corrections 
are sizable. On the lower panels we plot the ratios of the predictions of GGGH (bottom panel) 
and SW (middle panel) normalized to our results and find agreement between the various 
computations, except at the kinematic limits of the distributions.

In order to better quantify the level of agreement among the perturbative predictions, 
on the right hand panels of \fig{fig:omTphysCcoeff} and \fig{fig:CparphysCcoeff} we present 
the comparisons of the NNLO coefficients directly. We plot the distribution of the NNLO 
coefficient $O\, C(O)$ on the top panels, while the middle and bottom panels again show 
the ratios of the predictions of SW and GGGH normalized to our results. The narrow gray 
bands on the middle and lower panels show the numerical uncertainty of our computation due 
to Monte Carlo integrations. We observe a very good numerical convergence of our method 
at NNLO. Examining these plots, we see that the agreement is generally 
quite good between the predictions of SW and \colorfulmethod\ and reasonably good between 
GGGH and \colorfulmethod, with the precise comparison to GGGH being hampered by the somewhat 
large numerical uncertainties of those predictions. We also see that significant deviations 
are present for small and large values of the event shapes. For example the differences 
between \colorfulmethod\ and the two other predictions grow up to a factor of two for 
$\tau > 1/3$. However, in this region the contribution from the three-particle final 
state vanishes and the thrust distribution is determined by a four-jet final state. Hence, 
in this region $C(\tau)$ is given by NLO corrections to four-jet production, which have 
been known for a long time \cite{Signer:1996bf,Nagy:1997yn} and can also be computed by 
modern automated tools such as \amcatnlo\ \cite{Alwall:2014hca}. We find that our predictions 
are in complete agreement with those \amcatnlo\ for the thrust distribution for $\tau>1/3$, 
and with the computation of Ref.~\cite{Nagy:1997yn} for the C-parameter distribution for 
$C_{\mathrm{par}}>3/4$. For small values of the event shapes, we checked that our 
predictions are in agreement with the resummed computations obtained from SCET 
\cite{Becher:2008cf,Chien:2010kc,Hoang:2014wka} 
expanded to $\mathcal{O}(\alpha_{\rm s}^3)$.

Beside the standard event shape variables discussed above, we computed for the first 
time predictions at NNLO accuracy for oblateness, energy-energy correlation (EEC) 
\cite{DelDuca:2016csb} and jet cone energy fraction (JCEF) \cite{DelDuca:2016ily}. 
Here we present our results for jet cone energy fraction, which is defined as
\begin{align}
\frac{{\rm d}\Sigma_{\rm JCEF}}{{\rm d}\cos\chi} &=
\sum_{i}\int{\rm d}\sigma_{e^+\,e^-\to i+X}
\frac{E_i}{Q}\delta(\cos\chi + \cos\theta_{ij})
\delta\left(
\cos\chi - \frac{\vec{p}_i\cdot\vec{n}_T}{|\vec{p}_i|}
\right)
\,,
\end{align}
where $Q$ is the center of mass energy, $E_i$ is the energy of particle $i$ (in the center 
of mass frame), $\cos\theta_{ij}$ is the cosine of the angle between the three-momenta of 
particles $i$ and $j$ (also in the center of mass frame) and $\vec{n}_T$ is the thrust axis 
pointing from the heavy to the light jet mass hemisphere.
\begin{figure}[t]
\centering
\begin{tabular}{cc}
\includegraphics[width=0.47\textwidth]{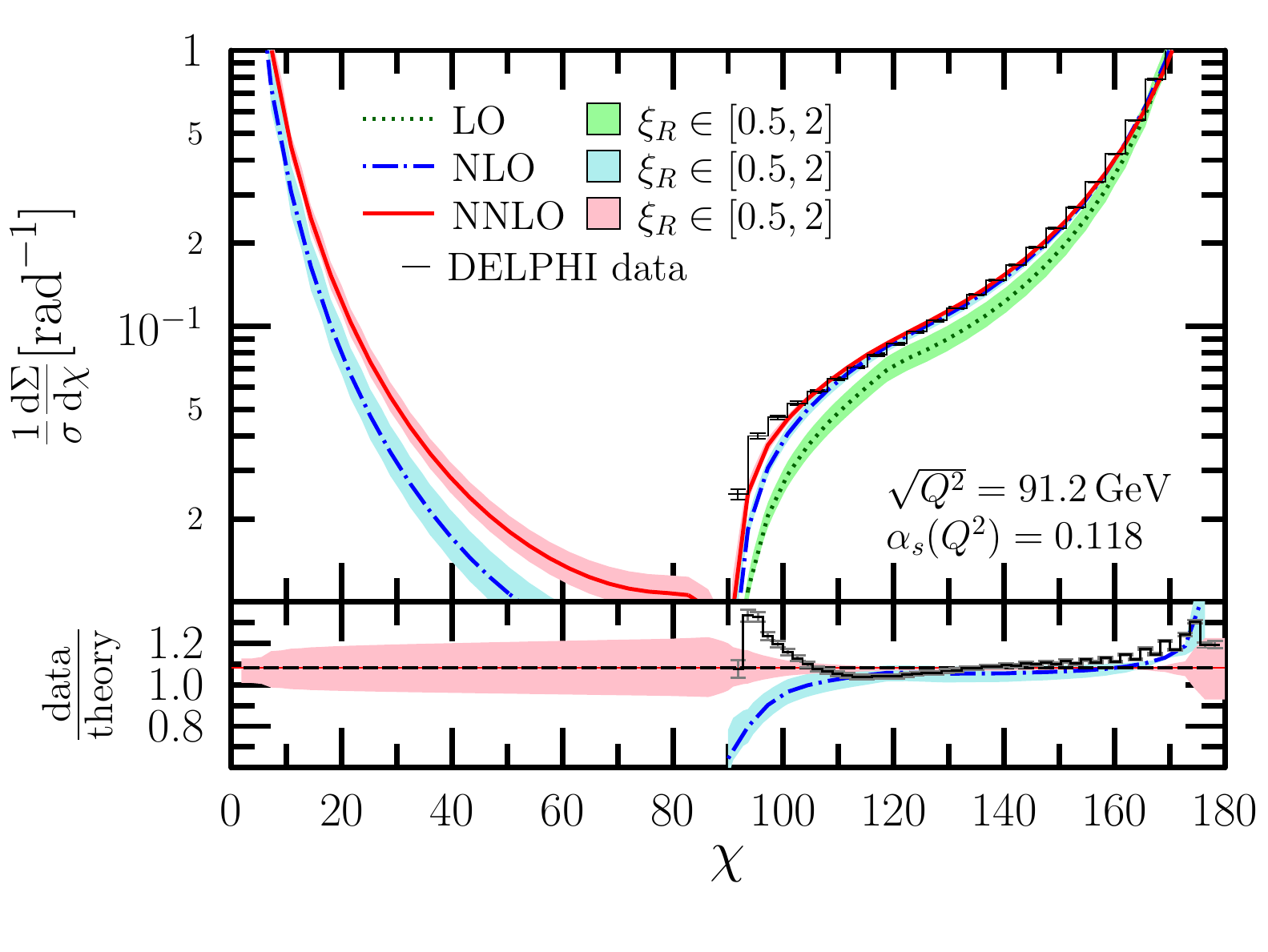} &
\includegraphics[width=0.47\textwidth]{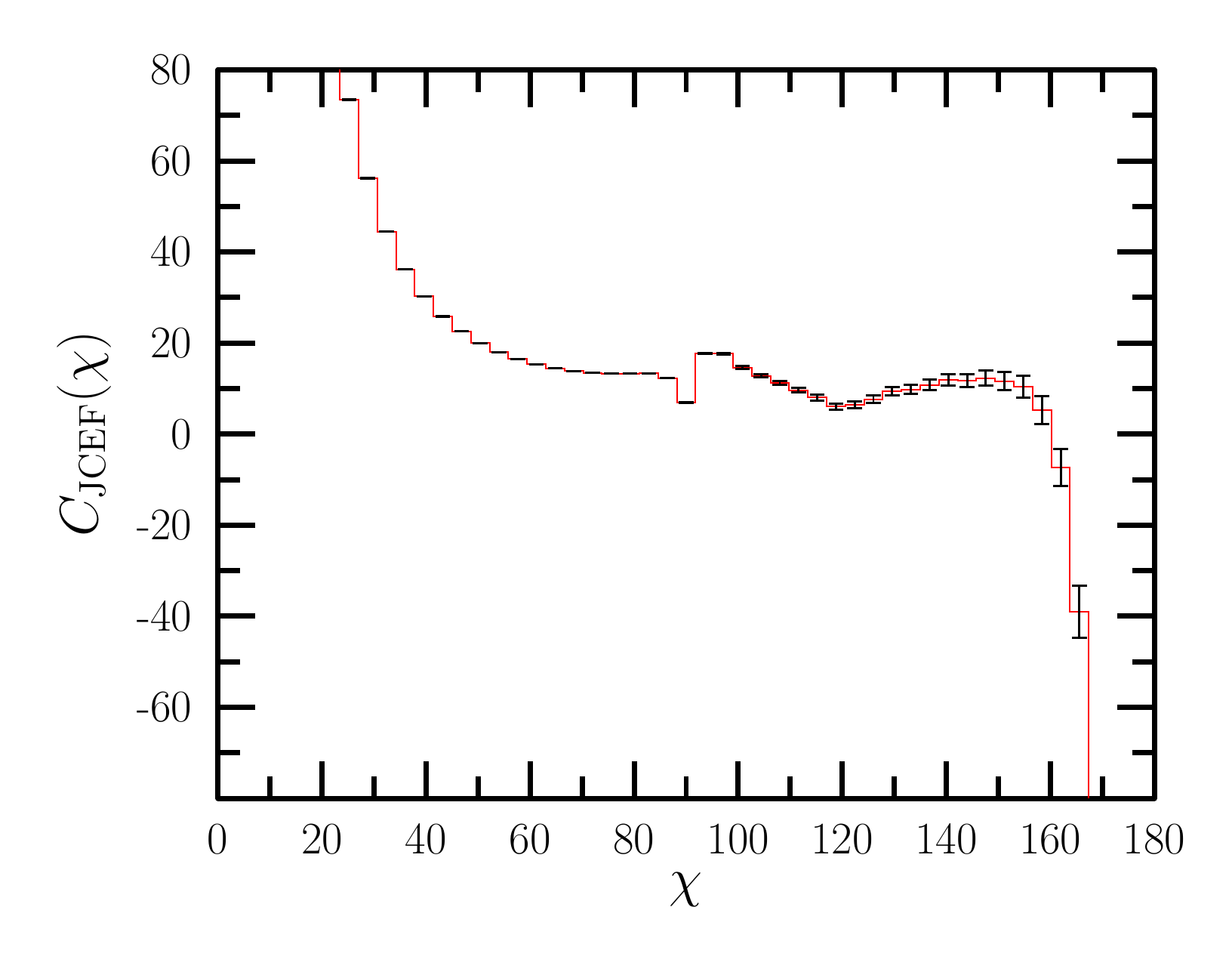} \\
\end{tabular}
\caption{{\label{fig:JCEF}}Left: physical predictions for jet cone energy fraction 
at LO, NLO and NNLO accuracy in QCD with bands representing scale uncertainty. 
Data measured by the DELPHI collaboration \cite{Abreu:2000ck} is also shown.
Right: the $C_{\mathrm{JCEF}}(\chi)$ NNLO coefficient of the jet cone energy fraction 
distribution. Error bars represent the numerical uncertainty coming from Monte Carlo 
integrations.}
\end{figure}
Our physical predictions for the jet cone energy fraction at LO, NLO and NNLO accuracy 
together with our prediction for the NNLO coefficient $C_{\mathrm{JCEF}}(\chi)$ are presented 
in \fig{fig:JCEF}. Our code displays a good numerical convergence also for these distributions. 
On the left hand panel showing the physical prediction, we have also included experimental 
data measured by the DELPHI collaboration. We observe that perturbative corrections are 
rather small over a wide range of angles. Hadronization corrections and detector corrections 
for this observable are also known to be quite small and indeed the perturbative result alone 
is seen to give a rather reasonable description of the data. Hence, jet cone energy fraction 
is a particularly simple and promising observable for the precise extraction of \as\ from 
data \cite{Abreu:2000ck}.

Finally, we turn to the computation of jet rates at NNLO accuracy. The production 
rate for $n$-jet events in electron-positron annihilation is given by the ratio of 
the $n$-jet cross section to the total hadronic cross section:
\begin{align}
R_{n}(y_{\mathrm{cut}}) = 
	\frac{\sigma_{n}(y_{\mathrm{cut}})}{\sigma_{\mathrm{tot}}}\,.
\end{align}
Here the $n$-jet cross section $\sigma_{n}(y_{\mathrm{cut}})$ must be defined 
using an infrared-safe jet clustering algorithm. One class of such algorithms are 
the exclusive sequential recombination algorithms. Here we focus on the Durham 
algorithm \cite{Catani:1991hj} for which the resolution variable is defined as
\begin{align}
y_{ij} = \frac{2\min(E_i^2,E_j^2)(1-\cos\theta_{ij})}{Q^2}
\end{align}
and recombination is performed in the E-scheme, i.e., the four-momenta of the objects 
to be combined are simply added.

\begin{figure}[t]
\centering
{
\setlength{\tabcolsep}{-0.5em}
\begin{tabular}{cc}
\includegraphics[width=0.55\textwidth]{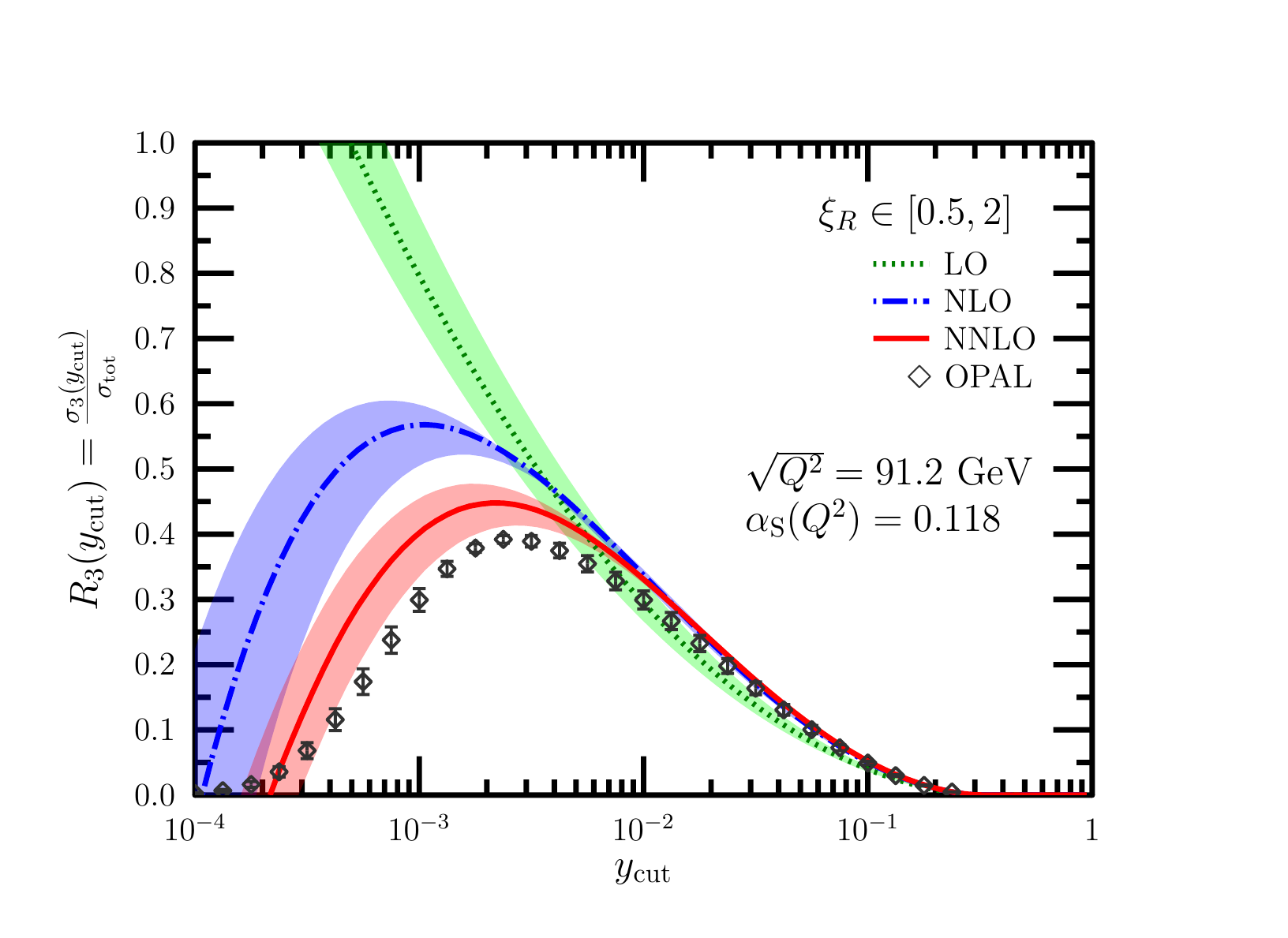} &
\includegraphics[width=0.55\textwidth]{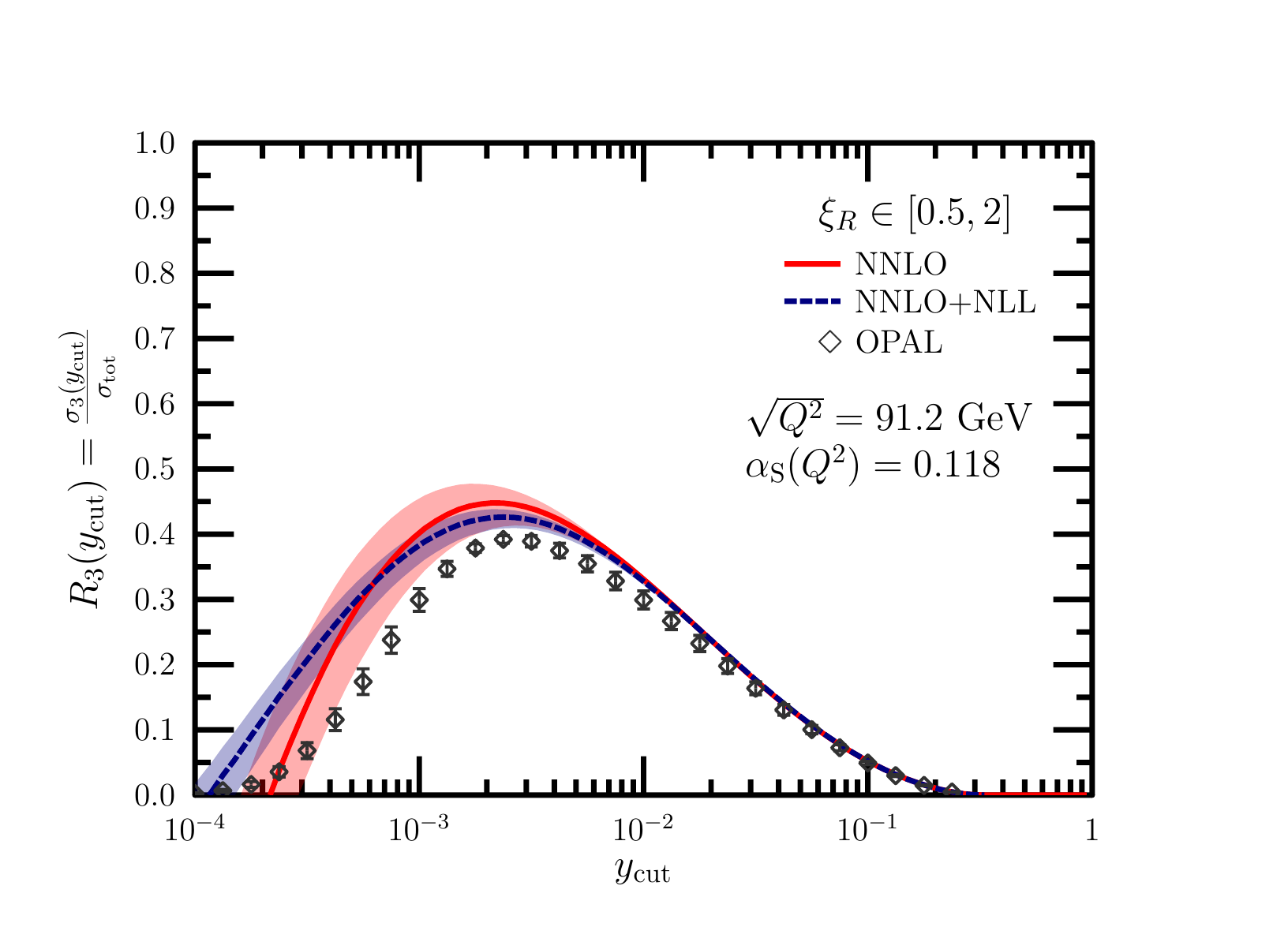} \\
\end{tabular}
}
\caption{{\label{fig:R3}}Left: preliminary physical predictions for the three-jet rate 
$R_{3}(y_{\mathrm{cut}})$ in the Durham clustering algorithm at LO, NLO and 
NNLO accuracy in QCD with bands representing scale uncertainty.
Data measured by the OPAL collaboration \cite{Pfeifenschneider:1999rz} is also shown.
Right: preliminary  predictions for the three-jet rate $R_{3}(y_{\mathrm{cut}})$ at NNLO 
and NNLO+NLL accuracy.}
\end{figure}

We present our preliminary physical predictions for $R_{3}(y_{\mathrm{cut}})$ 
in the Durham clustering algorithm in the left panel of \fig{fig:R3} at LO, NLO and NNLO 
accuracy, together with measured data from the OPAL collaboration. Although the inclusion 
of the NNLO corrections vastly improves the theoretical description of the measurement, 
a sizable difference remains between the NNLO prediction and the data for small values of 
the resolution parameter. We attribute this difference to missing parton shower (or 
resummation) and hadronization effects. In 
order to improve the situation, we match our perturbative prediction to the resummed result, 
which is known for $R_{3}(y_{\mathrm{cut}})$ in the two-jet limit (i.e., 
$y_{\mathrm{cut}} \to 0$) up to next-to-leading-logarithmic (NLL) accuracy \cite{Catani:1991hj}. 
(The distribution of the two-to-three jet transition variable $y_{23}$ in the Durham 
algorithm is known up to NNLL accuracy \cite{Banfi:2016zlc}.) This matching is performed 
as follows. We write the fixed order prediction for the three-jet rate at NNLO accuracy as
\begin{align}
R_3^{\mathrm{NNLO}}(y_{\mathrm{cut}}) = 
	\frac{\as}{2\pi} A_3(y_{\mathrm{cut}})
	+ \left(\frac{\as}{2\pi}\right)^2 B_3(y_{\mathrm{cut}})
	+ \left(\frac{\as}{2\pi}\right)^3 C_3(y_{\mathrm{cut}})\,.
\label{eq:R3NNLO}
\end{align}
The resummed prediction at NLL accuracy can be written in the following form,
\begin{align}
R_3^{\mathrm{NLL}}(y_{\mathrm{cut}}) = 
	2 [\Delta_q(Q)]^2 \int_{Q_0}^Q \mathrm{d} q\, \Gamma_q(Q,q) \Delta_g(q)\,,
\label{eq:R3NLL}
\end{align}
where $Q_0 = \sqrt{y_{\mathrm{cut}}} Q$ and the $\Delta_i(Q'')$ and $\Gamma_i(Q'',Q')$ 
functions are given explicitly in \cite{Catani:1991hj} up to NLL accuracy. 
When evaluating \eqn{eq:R3NLL} numerically, we use the one-loop running of \as\ in 
$\Delta_i(Q'')$ and $\Gamma_i(Q'',Q')$. In order to perform the matching, we expand 
\eqn{eq:R3NLL} in powers of \as\ up to and including $\mathcal{O}(\alpha_{\mathrm s}^3)$ 
terms:
\begin{align}
R_3^{\mathrm{NLL}}(y_{\mathrm{cut}}) &= 
	\frac{\as}{2\pi} A_3^{\mathrm{NLL}}(y_{\mathrm{cut}})
	+ \left(\frac{\as}{2\pi}\right)^2 B_3^{\mathrm{NLL}}(y_{\mathrm{cut}})
	+ \left(\frac{\as}{2\pi}\right)^3 C_3^{\mathrm{NLL}}(y_{\mathrm{cut}})
\notag\\&
	+ \mathcal{O}(\alpha_{\rm s}^4)\,.
\label{eq:R3NLLexp}
\end{align}
Our final expression at NNLO+NLL accuracy is then given by
\begin{align}
R_3^{\mathrm{NNLO+NLL}}(y_{\mathrm{cut}}) =
	R_3^{\mathrm{NLL}}(y_{\mathrm{cut}}) 
	&+ \frac{\as}{2\pi} 
		\Big[A_3(y_{\mathrm{cut}}) - A_3^{\mathrm{NLL}}(y_{\mathrm{cut}})\Big]
\notag\\&
	+ \left(\frac{\as}{2\pi}\right)^2
		\Big[B_3(y_{\mathrm{cut}}) - B_3^{\mathrm{NLL}}(y_{\mathrm{cut}})\Big]
\notag\\&
	+ \left(\frac{\as}{2\pi}\right)^3
		\Big[C_3(y_{\mathrm{cut}}) - C_3^{\mathrm{NLL}}(y_{\mathrm{cut}})\Big]\,.
\label{eq:R3NNLO+NLL}
\end{align}
The right hand panel of \fig{fig:R3} shows the preliminary results of this matching procedure. 
We indeed see a marked improvement of the theoretical description, together with a significant 
reduction in the relative scale uncertainty below $y_{\mathrm{cut}} \sim 10^{-2}$. Since 
jet rates computed using different jet algorithms can have different sensitivities to 
non-perturbative effects, it would be interesting to extend these results to other jet 
clustering algorithms as well.

%%%%%%%%%%%%%%%%%%%%%%%%%%%%%%%%%%%%%%%%%%%%%%%%%%%%%%%%%%%%
% Conclusions                                              %
%%%%%%%%%%%%%%%%%%%%%%%%%%%%%%%%%%%%%%%%%%%%%%%%%%%%%%%%%%%%

\section{Conclusions}

In this contribution we briefly outlined the \colorfulmethod\ subtraction method
for computing NNLO QCD corrections for processes with colorless initial states.
As a first application, the method was used to compute physical observables in 
three-jet production in electron-positron annihilation. After validating our numerical 
program by comparisons to existing computations, we presented NNLO QCD results for jet 
cone energy fraction, which has not been computed at NNLO accuracy before. We find that 
the perturbative corrections for this observable are rather small for a wide range of 
angles. This, together with the smallness of the hadronization and detector corrections 
makes the jet cone energy fraction a very promising observable for the precise extraction 
of the strong coupling from data. Finally, we presented preliminary results for the 
three-jet rate in the Durham algorithm at NNLO and NNLO+NLL accuracy.

%%%%%%%%%%%%%%%%%%%%%%%%%%%%%%%%%%%%%%%%%%%%%%%%%%%%%%%%%%%%
% References                                               %
%%%%%%%%%%%%%%%%%%%%%%%%%%%%%%%%%%%%%%%%%%%%%%%%%%%%%%%%%%%%

%%%%%%%%%%%%%%%%%%%%%%%%%%%%%%%%%%%%%%%%%%%%%%%%%%%%%%%%%%%%
%                                                          %
% End document                                             %
%                                                          %
%%%%%%%%%%%%%%%%%%%%%%%%%%%%%%%%%%%%%%%%%%%%%%%%%%%%%%%%%%%%

\end{document}